\documentstyle[aps,prl,epsf,floats]{revtex}

\draft

\begin{document}
\twocolumn[\hsize\textwidth\columnwidth\hsize\csname
@twocolumnfalse\endcsname

\title{New CMBR data and the cosmic neutrino background}

\author{Steen Hannestad}

\address{NORDITA, Blegdamsvej 17, DK-2100 Copenhagen, Denmark}

\date{\today}

\maketitle

\begin{abstract}
New precision Cosmic Microwave Background Radiation (CMBR)
anisotropy data are beginning to constrain physics beyond
the standard model, for example in the form of additional light 
particle species. These constraints are complementary to what
can be obtained from big bang nucleosynthesis (BBN) 
considerations because they apply to much later times.
We derive a constraint on the equivalent number of 
neutrino species, $N_\nu$,
from the presently
available data. Specifically we analyse two different CMBR data sets
to test the robustness of our results.
Analyzing only CMBR data yields an upper bound of $N_\nu \lesssim
17$ (95\% confidence). Adding large scale structure (LSS) data
from the PSC-z survey
tightens the upper bound slightly. However, the addition of
LSS data gives a non-trivial {\it lower} bound of $N_\nu \geq 
1.5/2.5$ (95\% confidence) for the two data sets.
This is the first independent indication of the presence of the
cosmological neutrino background which is predicted by the standard
model, and seen in big bang nucleosynthesis.
The value $N_\nu = 0$ is disfavoured at $3\sigma$ and $4\sigma$
for the two data sets respectively.
\end{abstract}

 \vskip1.9pc]


\section{Introduction}

Precision measurements of the anisotropy in the Cosmic Microwave 
Background Radiation (CMBR) have recently begun to probe cosmology
with high precision.
The measurements have delivered remarkably strong support
for inflation for the standard inflationary paradigm, i.e.\
a flat geometry and a initial fluctuation power spectrum which
is close to scale invariant.
Because of the high precision of the current measurement it is also
possible to probe various other parameters of the standard model.
In the present paper we study the current limits on the relativistic
energy density during recombination. The energy density is usually 
parameterized in terms of $N_\nu$, the equivalent number of standard
model neutrino species
\begin{equation}
N_\nu \equiv \frac{\rho}{\rho_{\nu_0}}.
\end{equation}
The standard model prediction is $N_\nu \simeq 3.04$, where the
0.04 comes from the fact that neutrinos are not completely decoupled
during the electron-positron annihilation in the early universe
\cite{decoupling}.

BBN considerations give the bound \cite{Lisi:1999ng}
\begin{equation}
2 \leq N_{\nu,{\rm BBN}} \leq 4 \,\,\, {\rm (95\% \,\, confidence)}
\label{eq:bbn}
\end{equation}
A bound on this parameter has been derived previously from CMBR data
\cite{Hannestad:2000hc,Mangano:2001mc,Esposito:2001sv,Orito:2000zb}. 
However, it was pointed out by Kneller
et al. \cite{Kneller:2001cd} that the bound is quite sensitive to 
assumptions about other cosmological parameters.

In the present paper we discuss in detail degeneracies between
$N_\nu$ and various other cosmological parameters, particularly 
the Hubble parameter, $H_0$.
Using two different compiled data sets we derive bounds on $N_\nu$.
We then go on to discuss the influence of including data from 
large scale structure surveys. It turns out that including LSS
data significantly narrows the allowed region for $N_\nu$.

Apart from providing a fairly robust upper limit on $N_\nu$, the
main result of the present paper is that the cosmic neutrino 
background has been detected at more than the $3 \sigma$ level
(i.e.\ $N_\nu =0$ is disallowed at the $99.7\%$ level).
The presence of the neutrino background is also detected by big bang
nucleosynthesis data (see Eq.~(\ref{eq:bbn})). 
However, this is the first independent 
cosmological detection.
The standard model value $N_\nu = 3$ is in all cases within
$2\sigma$ of the maximum of the likelihood function, so there
is no evidence for deviations from the standard model in the
present data.

\section{CMBR data analysis}

Several data sets of high precision are now publicly available.
In addition to the COBE \cite{Bennett:1996ce} data for small 
$l$ there are data from
BOOMERANG \cite{boom}, MAXIMA \cite{max}, DASI \cite{dasi} 
and several other experiments \cite{WTZ,qmask}.
Wang, Tegmark and Zaldarriaga \cite{WTZ} (hereafter WTZ) 
have compiled a combined data set
from all these available data, including calibration errors.
In order to test the robustness of our results, we do the analysis
of $N_\nu$ for two different data sets. The first is the combined
data of WTZ. The other consists of the COBE and Boomerang data,
including the quoted calibration error of Boomerang \cite{boom}. 
This second
data set avoids possible systematics in the compiled data set.
However, the final result for $N_\nu$ is practically the same
for both data sets.

The CMBR fluctuations are usually
described in terms of the power spectrum, which is again expressed in
terms of $C_l$ coefficients as $l(l+1)C_l$, where
\begin{equation}
C_l \equiv \langle |a_{lm}|^2\rangle.
\end{equation}
The $a_{lm}$ coefficients are given in terms of the actual temperature
fluctuations as
\begin{equation}
T(\theta,\phi) = \sum_{lm} a_{lm} Y_{lm} (\theta,\phi).
\end{equation}
Given a set of experimental measurements, the likelihood function is
\begin{equation}
{\cal L}(\Theta) \propto \exp \left( -\frac{1}{2} x^\dagger
[C(\Theta)^{-1}] x \right),
\end{equation}
where $\Theta = (\Omega, \Omega_b, H_0, n, \tau, \ldots)$ is a vector
describing the given point in parameter space. $x$ is a vector
containing all the data points and $C(\Theta)$ is the data covariance
matrix.  This applies when the errors are Gaussian. If we also assume
that the errors are uncorrelated, this can be reduced to the simple
expression, ${\cal L} \propto e^{-\chi^2/2}$, where
\begin{equation}
\chi^2 = \sum_{i=1}^{N_{\rm max}} \frac{(C_{l, {\rm obs}}-C_{l,{\rm
theory}})_i^2} {\sigma(C_l)_i^2},
\label{eq:chi2}
\end{equation} 
is a $\chi^2$-statistics and $N_{\rm max}$ is the number of power
spectrum data points \cite{oh}.  In the present letter we use
Eq.~(\ref{eq:chi2}) for calculating $\chi^2$.

The procedure is then to calculate the likelihood function over the
space of cosmological parameters. The 1D likelihood function for $N_\nu$
is obtained by keeping $N_\nu$ fixed and maximizing ${\cal L}$
over the remaining parameter space.

As free parameters in the likelihood analysis we use
$\Omega_m$, the matter density, $\Omega_b$, the baryon density,
$H_0$, the Hubble parameter, $n$, the scalar spectral index,
$\tau$, the optical depth to reionization, and $Q$ the overall
normalization of the data. When large scale structure constraints
are included we also use $b$, the normalization of the matter
power spectrum, as a free parameter. This means that we treat
$Q$ and $b$ as free and uncorrelated parameters. This is very
conservative and eliminates any possible systematics involved in
determining the bias parameter.
We constrain the analysis to flat ($\Omega_m + \Omega_\Lambda = 1$)
models, and we assume that the tensor mode contribution is 
negligible.
These assumptions are compatible with analyses of the present data
\cite{WTZ}, and relaxing them do not have a big effect on 
the final results.
For maximizing the likelihood function we use a simulated annealing
method, as described in Ref.~\cite{Hannestad:2000wx}.

\subsection{Priors}

As was shown by Kneller et al. \cite{Kneller:2001cd}, 
different priors can significantly
bias the derived confidence interval for $N_\nu$. We therefore 
test the effect
of different priors on the final result.
Table I shows the different priors used. In the ``weak'' prior 
the only important constraint is that $0.4 \leq h \leq 0.9$
($h \equiv H_0/(100 \,\, {\rm km} \, {\rm s}^{-1} \, {\rm Mpc}^{-1})$).
For the $H_0$+BBN prior we use the constraint 
$H_0 = 72 \pm 8 \,\, {\rm km} \, {\rm s}^{-1} \, {\rm Mpc}^{-1}$
from the HST Hubble key project \cite{freedman} (the constraint is added
assuming a Gaussian distribution) and the constraint $\Omega_b h^2
= 0.020 \pm 0.002$ from BBN \cite{Burles:2000zk}.
Finally, in the $H_0$+BBN+LSS case, we add data from the PSC-z survey
\cite{psc} to the data analysis.

The neutrino density is to some extent degenerate with other parameters,
particularly with the Hubble parameter. Increasing the Hubble parameter
allows for more neutrino species.
In the same manner, decreasing $n$ or $\Omega_b h^2$ 
allows for more relativistic
energy density. However, $N_\nu$ is only slightly degenerate with these 
parameters. 

In Fig.~1 we show the likelihood functions for the two different
data sets, assuming different priors. In the lower panels we show
values of other parameters for the best fits. From this, it is evident
that with only a weak prior on $H_0$, a large $N_\nu$ can be compensated
by increasing $H_0$. As soon as the HST Hubble key project prior on $H_0$
is added, the large values of $N_\nu$ are no longer allowed.

From this figure it can also be seen that there is very little 
degeneracy between $N_\nu$ and $n,\Omega_b h^2$. Furthermore, the
present data is entirely compatible with the BBN prior on $\Omega_b h^2$
(as can also be seen in Fig.~1). Therefore, adding the BBN prior
does not significantly change the analysis.

In Table II, the best fit values and the 95\% confidence limits
on $N_\nu$ are shown for the two data sets, for different priors.
Adding the prior $h = 0.72 \pm 0.08$ from the HST key project gives
a $2\sigma$ upper limit of $N_\nu \leq 17$ for the COBE+Boomerang
data set and $N_\nu \leq 17.5$ for the WTZ data set.

\subsection{LSS data}

Adding relativistic energy density also affects the matter power spectrum
because the growth factor on scales smaller than the horizon
is decreased (see e.g.~\cite{Dodelson:1994it}). 
In Figs.\ 2 and 3 we show the CMBR and matter power spectra for the best
fit models with different $N_\nu$ to the WTZ + PSC-z data sets. 

Changing $N_\nu$ clearly also changes the matter spectrum, especially
on scales of $0.01-0.1 h \,\, {\rm Mpc}^{-1}$
This fact can be used together with the CMBR data to constrain $N_\nu$.

\onecolumn

\begin{table}[t]
\caption{The different priors used in the analysis.}
\begin{tabular}{lccccccc}
prior type & $\Omega_m$ & $\Omega_b h^2$ & $h$ & $n$ & $\tau$ & $Q$ & $b$ \cr
\tableline 
``weak'' & $\Omega_b$-1 & 0.008 - 0.040 & 0.4-0.9 & 0.66-1.34 & 0-1 & free & - \cr
BBN + $H_0$ & $\Omega_b$-1 & $0.020 \pm 0.002$ & $0.72 \pm 0.08$ & 0.66-1.34 
& 0-1 & free & - \cr
BBN + $H_0$ + LSS & $\Omega_b$-1 & $0.020 \pm 0.002$ & $0.72 \pm 0.08$ & 0.66-1.34 
& 0-1 & free & free \cr
\end{tabular}
\end{table}

\begin{figure}[b]
\begin{center}
\hspace*{2cm}\epsfysize=15truecm\epsfbox{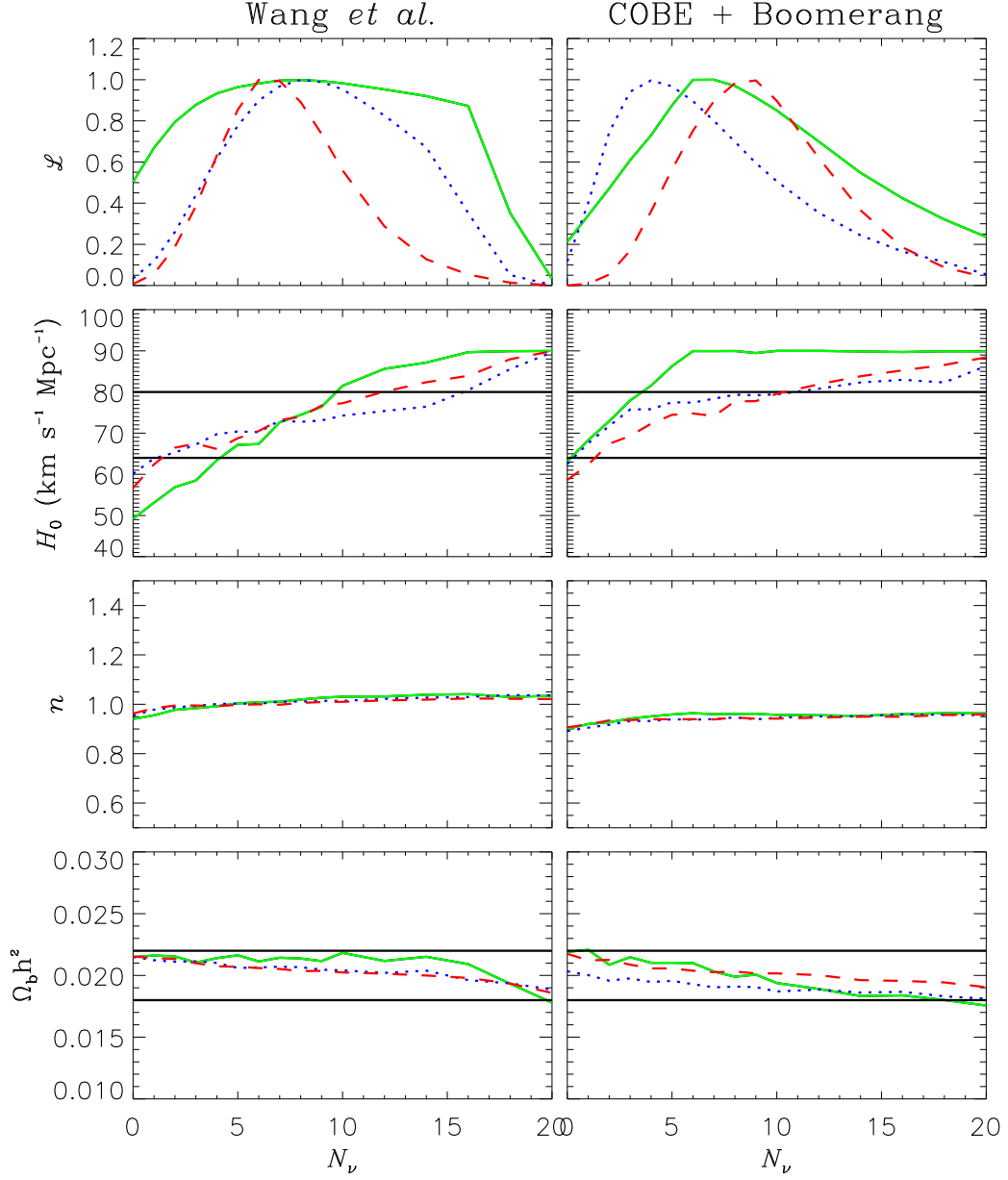}
\vspace{1.5truecm}
\end{center}
\caption{The top panels show the likelihood functions for the
two different data sets, including different priors. The full
lines are for the ``weak'' prior, the dotted for the $H_0+\Omega_b h^2$
prior, and the dashed for the $H_0+\Omega_b h^2$+LSS prior.
The lower panels show values of $H_0,\Omega_b h^2$ and $n$ for the
best fit models. Horizontal full lines show the HST key project
limit on $H_0$ and the BBN prior on $\Omega_b h^2$.}
\label{fig1}
\end{figure}

\twocolumn

Note that on even smaller scales, data from the Ly-$\alpha$ forest
\cite{Croft:2000hs}
can also be used. However, the very smallest scales are not so sensitive
to $N_\nu$ because the shape of the power spectrum is not changed by
adding radiation. The normalization is changed, but since we treat
the overall normalization of the power spectrum as a free parameter,
this will not have any effect. We therefore only use data from the
PSC-z survey to give the LSS constraints.
Adding the LSS data again tightens the constraint. 
The likelihood functions and best fit parameter values when LSS data
is included can also be seen in Fig.~1.
The $2\sigma$ upper
limits are now $N_\nu \leq 17$ for the COBE+Boomerang
data set and $N_\nu \leq 14$ for the WTZ data set.

For the WTZ data the upper bound is lowered from 17.5 to 14 by adding
LSS data. The effect can be seen in Figs.~2-3, for the model with 
$N_\nu=14$. Although this model can provide a very good fit to CMBR
data, the shape of the matter spectrum becomes too shallow to obtain
a decent fit.

Very interestingly there is now also a non-trivial lower bound on
$N_\nu$ which is $N_\nu \geq 2.5$ for for the COBE+Boomerang
data set and $N_\nu \geq 1.5$ for the WTZ data set.
$N_\nu =0$ is inconsistent with the data at roughly $4\sigma$ for 
COBE+Boomerang and $3\sigma$ for WTZ.
Indeed this result can be taken as the first real detection of the 
cosmological neutrino background at late epochs. From BBN considerations
one already has the result $N_\nu \gtrsim 2$ \cite{Lisi:1999ng}. 
However, there is now
an independent confirmation of the presence of relativistic energy
density other than photons. Since the CMBR is only sensitive to 
radiation and not to the specific content, it is impossible to tell
whether this radiation stems from the neutrinos as predicted by the
standard model, or from other light particles. However, the standard
result $N_\nu = 3.04$ is in all cases compatible with the data at the 
$2\sigma$ level.

The incompatibility of $N_\nu=0$ with data can also be seen in Figs.~2-3.
Although a good fit to LSS data can be obtained, the fit to CMBR data
is very poor. This is mainly because the first peak is too low due
to the absence of the early integrated Sachs-Wolfe effect
\cite{Hannestad:1998cv}.
\begin{table}
\caption{Best fit values and $2\sigma$ (95\%) limits on $N_\nu$ for
different priors and the two different data sets.}
\begin{tabular}{lcc}
prior type & WTZ & COBE+Boomerang \cr
\tableline 
``weak'' & $8^{+11}_{-8}$ & $7^{+17}_{-7}$ \cr
BBN + $H_0$ & $8^{+9.5}_{-7}$ & $4^{+13}_{-4}$ \cr
BBN + $H_0$ + LSS & $6^{+8}_{-4.5}$ & $9^{+8}_{-6.5}$  \cr
\end{tabular}
\end{table}

\begin{figure}[h]
\begin{center}
\epsfysize=7truecm\epsfbox{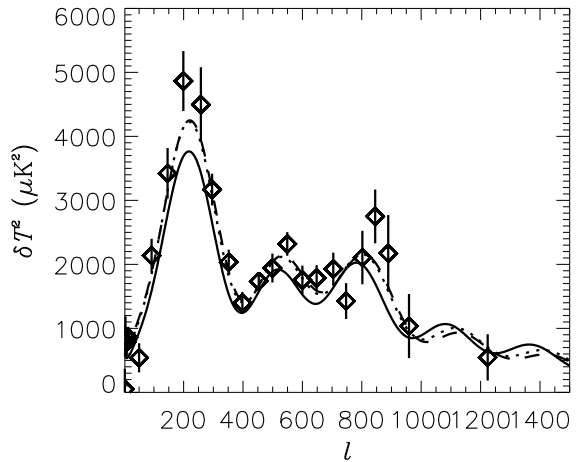}
\vspace{0truecm}
\end{center}
\caption{CMBR power spectra for the best fits to the WTZ+LSS data,
for $N_\nu = 0$ (full line), 7 (dashed line) and 14 (dotted line).
The data points are from the WTZ compiled data set.}
\label{fig2}
\end{figure}

\begin{figure}[h]
\begin{center}
\epsfysize=7truecm\epsfbox{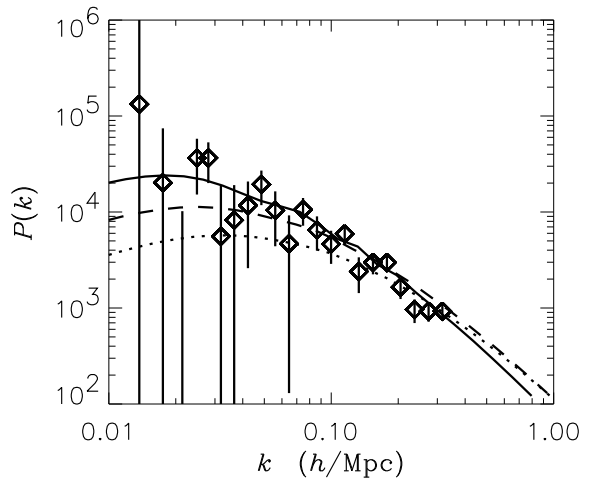}
\vspace{0truecm}
\end{center}
\caption{matter power spectra for the best fits to the WTZ+LSS data,
for $N_\nu = 0$ (full line), 7 (dashed line) and 14 (dotted line). 
The normalization is arbitrary and the data points are from the
PSC-z survey.}
\label{fig3}
\end{figure}

\section{discussion}

We have calculated bounds on the relativistic energy density present
during recombination from the present CMBR and LSS data.
The new data give a robust upper bound of $N_\nu \leq 17$, but, perhaps
more interesting, also give a {\it lower} bound of $N_\nu \geq 1.5/2.5$
for the two different data sets analysed.
Both bounds are interesting and non-trivial. Although the upper bound
is much weaker than the bound $N_\nu \leq 4$ found from BBN, it applies
to any type of relativistic energy density. The BBN bound can be avoided
by putting some of the extra energy density in electron neutrinos,
because these directly influence the neutron-proton conversion processes
prior to BBN 
\cite{Mangano:2001mc,Esposito:2001sv,Orito:2000zb,Lesgourgues:2000eq}.
The CMBR directly probes the energy density and is insensitive to 
flavour. The two constraints should therefore be seen as complementary.
Furthermore, if there are massive particles decaying after 
BBN, but prior to recombination, the light decay products will add to
the radiation density during recombination, but not during BBN.
This is the case in some decaying neutrino scenarios 
\cite{Dodelson:1994it,Hansen:2000td}, 
as well as in some scenarios with large extra dimensions 
\cite{Abazajian:2000hw}.

The lower limit on $N_\nu$ is highly interesting because it provides
the first strong indication of relativistic energy density other than
photons around the epoch of recombination. 
The value $N_\nu = 0$ is strongly disfavoured by the data, deviating
from the best fit by $3\sigma$ for the WTZ+LSS data set and 
$4\sigma$ for the COBE+Boomerang+LSS data.

Finally, although a central value
higher than $N_\nu =3$ seems to be preferred in the data, the standard
model value $N_\nu=3.04$ is compatible with the present data at the
$2\sigma$ level. This means that there is no significant indication
of non-standard physics contributing to $N_\nu$ at the recombination
epoch.


\begin{references}

\bibitem{decoupling}
D.~A.~Dicus, E.~W.~Kolb, A.~M.~Gleeson, E.~C.~Sudarshan, V.~L.~Teplitz and M.~S.~Turner,
Phys.\ Rev.\ D {\bf 26}, 2694 (1982),
N.~C.~Rana and B.~Mitra,
Phys.\ Rev.\ D {\bf 44}, 393 (1991),
S.~Dodelson and M.~S.~Turner,
Phys.\ Rev.\ D {\bf 46}, 3372 (1992),
A.~D.~Dolgov and M.~Fukugita,
Phys.\ Rev.\ D {\bf 46}, 5378 (1992),
S.~Hannestad and J.~Madsen,
Phys.\ Rev.\ D {\bf 52}, 1764 (1995)
[astro-ph/9506015],
A.~D.~Dolgov, S.~H.~Hansen and D.~V.~Semikoz,
Nucl.\ Phys.\ B {\bf 503}, 426 (1997)
[hep-ph/9703315],
N.~Y.~Gnedin and O.~Y.~Gnedin,
astro-ph/9712199.


\bibitem{Lisi:1999ng}
E.~Lisi, S.~Sarkar and F.~L.~Villante,
Phys.\ Rev.\ D {\bf 59}, 123520 (1999)
[hep-ph/9901404].


\bibitem{Hannestad:2000hc}
S.~Hannestad,
Phys.\ Rev.\ Lett.\  {\bf 85}, 4203 (2000)
[astro-ph/0005018].

\bibitem{Mangano:2001mc}
G.~Mangano, A.~Melchiorri and O.~Pisanti,
Nucl.\ Phys.\ Proc.\ Suppl.\  {\bf 100}, 369 (2001)
[astro-ph/0012291].

\bibitem{Esposito:2001sv}
S.~Esposito, G.~Mangano, A.~Melchiorri, G.~Miele and O.~Pisanti,
Phys.\ Rev.\ D {\bf 63}, 043004 (2001)
[astro-ph/0007419].


\bibitem{Orito:2000zb}
M.~Orito, T.~Kajino, G.~J.~Mathews and R.~N.~Boyd,
astro-ph/0005446.


\bibitem{Kneller:2001cd}
J.~P.~Kneller, R.~J.~Scherrer, G.~Steigman and T.~P.~Walker,
astro-ph/0101386.

\bibitem{Bennett:1996ce}
C.~L.~Bennett {\it et al.},
Astrophys.\ J.\  {\bf 464} (1996) L1
[astro-ph/9601067].


\bibitem{boom}C.~B.~Netterfield {\it et al.}, astro-ph/0104460.


\bibitem{max}A.~T.~Lee {\it et al.}, astro-ph/0104459.

\bibitem{dasi}N.~W.~Halverson {\it et al.}, astro-ph/0104489.

\bibitem{WTZ}X.~Wang, M.~Tegmark and M.~Zaldarriaga,
astro-ph/0105091 (WTZ).

\bibitem{qmask}X.~Yu {\it et al.}, astro-ph/0010552.

\bibitem{oh}S.~P.~Oh, D.~N.~Spergel and G.~Hinshaw, Astrophys.\ J.\
{\bf 510}, 551 (1999).


\bibitem{Hannestad:2000wx}
S.~Hannestad,
Phys.\ Rev.\ D {\bf 61}, 023002 (2000).


\bibitem{freedman}W.~L.~Freedman {\it et al.}, astro-ph/0012376.

\bibitem{Burles:2000zk}
S.~Burles, K.~M.~Nollett and M.~S.~Turner,
astro-ph/0010171.


\bibitem{psc}
M.~Tegmark, M.~Zaldarriaga and A.~J.~Hamilton,
Phys.\ Rev.\ D {\bf 63}, 043007 (2001)
[astro-ph/0008167],
A.~J.~S.~Hamilton, M.~Tegmark and N.~Padmanabhan, 
Mon.\ Not.\ R.\ Astron.\ Soc.\ {\bf 317}, L23 (2000);
W.~Saunders {\it et al.}, 
Mon.\ Not.\ R.\ Astron.\ Soc.\ {\bf 317}, 55 (2000).


\bibitem{Dodelson:1994it}
S.~Dodelson, G.~Gyuk and M.~S.~Turner,
Phys.\ Rev.\ Lett.\  {\bf 72}, 3754 (1994)
[astro-ph/9402028].


\bibitem{Croft:2000hs}
R.~A.~Croft {\it et al.},
astro-ph/0012324.


\bibitem{Hannestad:1998cv}
S.~Hannestad,
Phys.\ Lett.\ B {\bf 431}, 363 (1998)
[astro-ph/9804075].


\bibitem{Lesgourgues:2000eq}
J.~Lesgourgues and M.~Peloso,
Phys.\ Rev.\ D {\bf 62}, 081301 (2000)
[astro-ph/0004412].

\bibitem{Hansen:2000td}
S.~H.~Hansen and F.~L.~Villante,
Phys.\ Lett.\ B {\bf 486}, 1 (2000)
[astro-ph/0005114].


\bibitem{Abazajian:2000hw}
K.~Abazajian, G.~M.~Fuller and M.~Patel,
hep-ph/0011048.

\end{references}
\end{document}